\documentclass[twocolumn,showpacs,preprintnumbers,amsmath,amssymb,floatfix]{revtex4-1}

\usepackage{latexsym}
\usepackage{amsmath}
\usepackage{epsfig}
\input{epsf.sty}  

\usepackage{epstopdf}
\usepackage{graphicx}
\usepackage{dcolumn}
\usepackage{bm}
\usepackage{subfigure}
\usepackage{enumerate}

\begin{document}

\preprint{version/10-12-17}

\title{Static dilaton space-time parameters from frequency shifts of photons  
emitted by geodesic particles.}

\author{
Susana Valdez-Alvarado$^{1}$,
Ricardo Becerril$^{2}$,
and
Francisco Astorga$^{2}$,
}

\affiliation{
$^{1}$Facultad de Ciencias de la Universidad Aut\'onoma del Estado de
M\'exico, Instituto Literario No. 100, C.P. 50000, Toluca, Estado 
de M\'exico.
$^{2}$Instituto de F\'isica y Matem\'aticas, Universidad
Michoacana de San Nicol\'as de Hidalgo. Edif. C-3, 58040 Morelia, 
Michoac\'an, M\'exico. \\
}
\date{\today}

\begin{abstract}

The mass parameter of dilaton space-times
is obtained as a function of the redshift-blueshift ($z_{red},z_{blue}$) 
of photons emitted by particles orbiting in circular motion around 
these objects and their corresponding radii. Particularly,
we work with the generalized Chatterjee and Gibbons-Maeda 
space-times. Both of them become the Schwarzschild black hole in certain
limit of one of their parameters. Bounds for the values of these 
frequency shifts, that may be observed for these metrics, are also determined.
 
\end{abstract}

\maketitle

\section{Introduction}\label{sec:int}

Motivated by an increasing amount of observational evidence that 
in the center of several galaxies there is a black hole \cite{evidence},
Herrera-Nucamendi (hereafter refered to as HN) 
\cite{ulises} developed recently, a theoretical approach 
to determine the mass and rotation parameters of a Kerr black hole in terms
of the redshift-blueshift $z_{red}$, $z_{blue}$ 
of the photons emitted by particles orbiting around the black hole and 
the radii of their trajectories. 
In HN, an explicit expression for the rotation parameter 
$a=a(r_c,z_{red},z_{blue},M)$ was found, where $r_c$ is the radius of 
circular orbits; however, the mass parameter $M$ might only be found 
by solving an eighth order polynomial. These circular 
orbits are required to be bounded and stable. In this context, 
given a set of observational data $\{z_{red},z_{blue},r_c\}$, that is,
a set of red and blue shifts emitted by particles orbiting 
a Kerr black hole at different radii, what one would like to know is the
mass and rotation parameters in terms of that data set. 
A detailed analysis of how this can be accomplished was recently 
performed in \cite{us}, where the mass of the black hole  
for SgrA$^{*}$ and its corresponding 
angular momentum (recently estimated
\cite{estimates}: $M \sim 2.72 \times 10^6 M_{\odot}$
and $a \sim 0.9939 M$) were employed. 
In that work, it was also found the mass parameter of axialsymmetric 
non-rotating compact objects such as Schwarzschild and Reissner-Nordstrom 
black holes in terms of the red-blue shift of light 
and the orbiting radius of emitting particles. 

In this analysis, we carry out this sort of study on
Einstein-Maxwell-Dilaton (EMD) spacetimes. Dilaton fields coupled 
to Einstein-Maxwell fields appear naturally in the low energy limit 
of string theory and as a result of dimensional reduction of the 
5-dimensional Kaluza-Klein theory. As a candidate for dark matter, 
dilaton fields have been employed to explain the large scale structure 
of the universe \cite{LS} and at galactic 
level, to explain the rotation curves \cite{RC}. A dilatonic compact object
has also been studied as a gravitational lens \cite{Lens}. More recently, 
the properties and dynamics of black holes within the EMD theory has been
considered \cite{Dynamics}. In this paper, we apply the HN 
approach to some static exact solution of the EMD theory, namely, 
the Chatterjee class of solutions \cite{MaciasMatos} and
the Gibbons-Maeda spacetime \cite{GM}. 
Both solutions have the Schwarzschild black hole as a special case in certain 
limit of one of their parameters, which amounts to make the dilaton field
vanish. In \cite{us}, an explicit formula of $M=M(r_c,z)$ for the Schwarzscild
black hole was found, and it turned out that $z$ is bounded as
$|z|<1/\sqrt{2}$.  
Hence, a study of the influence of the dilaton field upon attainment
of $M=M(r_c,z)$ and the bounds of $z$ is carried out in this analysis.
We provide a brief summary of the HN theoretical scheme in section II, 
and in section III we deal with the non-rotating examples above mentioned.
 
\section{H-N Theoretical Approach}
\label{review}

Starting with a rotating axialsymmetric space-time
in spherical coordinates $(x^{\mu})=(t,r,\theta,\phi)$, 
the geodesic equations for a massive 
particle stem from the Lagrangian given by 

\begin{equation}
\mathcal{L}= \frac{1}{2} \left (
g_{tt} \dot{t}^2+2 g_{t\phi}\dot{t} \dot{\phi} + g_{rr} \dot{r}^2 + 
g_{\theta \theta} \dot{\theta}^2+ g_{\phi \phi} \dot{\phi}^2 \right )
\label{Lagrangian}
\end{equation}
 
The photons emitted by this massive particle move along null geodesics
whose equations come also from the same Lagrangian.
We assume that the space time is endowed 
with two Killing vectors $\xi=(1,0,0,0)$, $\psi=(0,0,0,1)$, 
hence the metric and the Lagrangian 
depend only on the variables $r$ and $\theta$
; thus, there are two quantities 
that are conserved along the geodesics of the massive particles

\begin{eqnarray}
p_t&=& \frac{\partial \mathcal{L}}{\partial \dot{t}} = 
g_{tt} \dot{t} + g_{t \phi} \dot{\phi}= 
g_{tt} U^t + g_{t \phi} U^{\phi}= -E, \nonumber \\
p_{\phi} &=& \frac{\partial \mathcal{L}}{\partial \dot{\phi}} = 
g_{t \phi} \dot{t} + g_{\phi \phi} \dot{\phi}= 
g_{t \phi} U^t + g_{\phi \phi} U^{\phi}= L,
\label{constants}
\end{eqnarray}

\noindent where $U^{\mu}=\frac{d x^{\mu}}{d \tau}$ is the 4-velocity 
and $\tau$ is the proper
time. This 4-velocity is normalized to unity rendering

\begin{eqnarray}
-1 &=& g_{tt} (U^t)^2 + g_{rr} (U^r)^2 + g_{\theta \theta} (U^{\theta})^2+
g_{\phi \phi} (U^{\phi})^2 \nonumber \\
&& + g_{t \phi} U^t U^{\phi}.
\label{condition}
\end{eqnarray}

\noindent 
Using (\ref{constants}), two of these  4-velocity components can 
be written in terms of $g_{\mu \nu}$, $E$ and $L$ as 

\begin{equation}
U^t = \frac{g_{\phi \phi} E + g_{t \phi} L}{g_{t\phi}^2 -g_{tt} g_{\phi \phi}}
\quad , \quad
U^{\phi} = -\frac{g_{t \phi} E + g_{tt } L}{g_{t\phi}^2 -g_{tt} g_{\phi \phi}}.
\label{us}
\end{equation}

\noindent 
then (\ref{condition}) yields

\begin{equation}
g_{rr} \left ( U^r \right )^2 + V_{eff} =0,
\label{veff}
\end{equation}

\noindent 
where $V_{eff}$ is an effective potential given by

\begin{equation}
V_{eff}=
1+ g_{\theta \theta} \left ( U^{\theta} \right )^2 -
\frac{E^2 g_{\phi \phi}+ L^2 g_{tt} + 2 E L g_{t \phi}}
{g_{t \phi}^2-g_{tt}g_{\phi \phi}}.
\label{veffexplicit}
\end{equation}

\noindent 
Our aim is to get the parameters of an axialsymmetric 
space-time in terms of the red and blue shifts 
$z_{red}$ and $z_{blue}$ of light
emitted by massive particles orbiting around a compact object.
These photons have 4-momentum $k^{\mu}=(k^t,k^r,k^{\theta},k^{\phi})$ 
and move along null geodesics $k_{\mu} k^{\mu}=0$. 
Using the same Lagrangian (\ref{Lagrangian})
one obtains two conserved quantities:
$-E_{\gamma}= g_{tt} k^t + g_{t\phi} k^{\phi}$ and
$L_{\gamma} =g_{\phi t} k^t + g_{\phi \phi} k^{\phi}$. By inverting these two
expressions one can write $k^t$ and $k^{\phi}$ in terms of $g_{\mu \nu}$, 
$E_{\gamma}$ and $L_{\gamma}$.

The frequency shift $z$ associated to the emission and detection of 
photons is defined as

\begin{equation}
1+z= \frac{\omega_e}{\omega_d}=
\frac{- k_{\mu} U^{\mu} |_e }{- k_{\nu} U^{\nu} |_d}.
\label{zzz}
\end{equation}

\noindent where $\omega_e$ is the frequency emitted by an observer 
moving with the massive particle at the emission point $e$ and $\omega_d$
the frequency detected by an observer far away from the source
of emission, $U^{\mu}_e$ and $U^{\mu}_d$ are the 4-velocity of the emitter and
detector respectively. If the detector is located very far away from
the source ($r \to \infty$) then 
$U^{\mu}_d=(1,0,0,0)$ since $U^r_d,U^{\theta}_d,U^{\phi}_d \to 0$,
whereas $U^t=E=1$. The frequency $\omega_e= -k_{\mu}U^{\mu}|_e$ is 
explicitly given by \\

$\omega_e = \left ( E_{\gamma}U^t-L_{\gamma}U^{\phi}-g_{rr} U^r k^r-
g_{\theta \theta} U^{\theta} k^{\theta} \right)|_e $,\\

\noindent 
with a similar expression for $\omega_d$; hence (\ref{zzz}) becomes

\begin{equation}
1+z=
\frac{\left ( E_{\gamma}U^t-L_{\gamma}U^{\phi}-g_{rr} U^r k^r-
g_{\theta \theta} U^{\theta} k^{\theta} \right)|_e}
{ \left ( E_{\gamma} U^t - L_{\gamma} U^{\phi}-g_{rr} U^r k^r-
g_{\theta \theta} U^{\theta} k^{\theta} \right)|_d }.
\label{oneplusz}
\end{equation}

\noindent This is a general expression for the red-blue shifts of light
emitted by massive particles that are orbiting around a compact object
measured by a distant observer. We shall study the particular case of
circular ($U^r=0$) and equatorial ($U^{\theta}$=0) motion which 
simplify the expression (\ref{oneplusz}) to

\begin{equation}
1+z=
\frac{\left ( E_{\gamma}U^t-L_{\gamma}U^{\phi} \right)|_e}
{ \left ( E_{\gamma} U^t - L_{\gamma} U^{\phi} \right)|_d }
= \frac{U^t_e-b_eU^{\phi}_e}{U^t_d - b_d U^{\phi}}
\label{simple}
\end{equation}

\noindent where the apparent impact parameter $b=L_{\gamma}/E_{\gamma}$
of photons was introduced. Since it is defined in terms of quantities that are 
conserved all the way from the point of emission to the point of 
detection, one has that $b_e=b_d$. In addition, the 
kinematic redshift-blueshift of photons $z_{kin}=z-z_c$  was considered;
here $z_c$ is the redshift corresponding to a photon emitted by a static 
particle located at $b=0$

\begin{equation}
1+z_c= \frac{U^t_e}{U^t_d}.
\label{zc}
\end{equation}

\noindent The kinematic redshift $z_{kin} = (1+z) - (1+z_c)$ can 
then be written as 

\begin{equation}
z_{kin} =
\frac{(U^t-bU^{\phi}-\frac{1}{E_{\gamma}}g_{rr}U^rk^r-\frac{1}{E_{\gamma}}
  g_{\theta \theta} U^{\theta}k^{\theta})|_e}
{(U^t-bU^{\phi}-\frac{1}{E_{\gamma}}g_{rr}U^rk^r-\frac{1}{E_{\gamma}}
  g_{\theta \theta} U^{\theta}k^{\theta})|_d}-\frac{U^t_e}{U^t_d}
\label{zcine}
\end{equation}

\noindent
The analysis may be performed with either $z_{kin}$ using (\ref{zcine}) 
or with $z$ using (\ref{oneplusz}). We work with $z_{kin}$ in this paper. 
The general expression (\ref{zcine}) acquires a rather simply form for
circular orbits ($U^r=0$) in the equatorial plane ($U^{\theta}=0$)

\begin{equation}
z_{kin}= \frac{U^t U_d^{\phi} b_d- U^t_d U^{\phi}_e b_e}
{U^t_d(U^t_d-b_d U^{\phi}_d)}.
\label{zkin}
\end{equation}

In (\ref{zkin}) we still need to take into account light 
bending due to gravitational field, that is, to find
$b$ as a function of the location of the emitter $r$: 
$b=b(r)$. The criteria employed in \cite{ulises} to 
construct this mapping was choosing the maximum value of $z$ at a fixed
distance from the observed center of the source (at a fixed
$b$). From $k_{\mu} k^{\mu}=0$ with $k^r=k^{\theta}=0$ one arrives at

\begin{equation}
b_{\pm}= \frac{-g_{t \phi} \pm \sqrt{g_{t \phi}^2-g_{tt}g_{\phi \phi}}}{g_{tt}},
\label{bmm}
\end{equation}

\noindent $b_{\pm}$ can be evaluated at the emitter or detector
position. Since in general there are two different values of
$b_{\pm}$, there will be two different values of $z$ of photons
emitted by a receding ($z_1$) or an approaching object ($z_2$) with
respect to a distant observer. These kinematic shifts of photons emitted
either side of the central value $b=0$ read

\begin{equation}
z_1=\frac{U^t_e U^{\phi}_d b_{d_{-}}-U^t_d U^{\phi}_e b_{e_{-}}}
{U^t_d(U^t_d-U^{\phi}_d b_{d_{-}})},
\label{z1}
\end{equation}

\begin{equation}
z_2=\frac{U^t_e U^{\phi}_d b_{d_{+}}-U^t_d U^{\phi}_e b_{e_{+}}}
{U^t_d(U^t_d-U^{\phi}_d b_{d_{+}})}.
\label{z2}
\end{equation}

For the case of static space-times, that is, for
$g_{t \phi}=0$, the apparent impact parameter becomes
$b_{\pm}= \pm \sqrt{-g_{\phi \phi}/g_{t t}}$ 
and the effective potential (\ref{veffexplicit}) acquires a 
rather simple form

\begin{equation}
V_{eff} = 1 + \frac{E^2}{g_{t t}}+\frac{L^2}{g_{\phi \phi}}.
\label{potentialreduce}
\end{equation}

\noindent 
For circular orbits, $V_{eff}$ and its derivative
$\frac{d V_{eff}}{dr}$ vanish. 
From these two conditions one finds two general expressions 
for the constants of motion $E^2$ and $L^2$ valid for any non-rotating
axialsymmetric space-time

\begin{equation}
E^2=-\frac{g_{tt}^2 g_{\phi \phi}^{\prime}}
{g_{tt} g_{\phi\phi}^{\prime}-g_{tt}^{\prime} g_{\phi \phi}},
\quad L^2=\frac{g_{\phi \phi}^2 g_{tt}^{\prime}}
{g_{tt} g_{\phi\phi}^{\prime} - g_{tt}^{\prime}g_{\phi \phi} },
\label{E2L2}
\end{equation}

\noindent 
where primes denote derivative with respect to $r$.
Stability of these circular orbits requires 
$V_{eff}^{\prime \prime}>0$. If we use (\ref{E2L2}), a general expression for 
$V_{eff}^{\prime \prime}$ in terms of $g_{\mu \nu}$ and its derivatives is found

\begin{equation}
V_{eff}^{\prime \prime} = 
\frac{g_{\phi \phi}^{\prime} g_{tt}^{\prime\prime}
 - g_{tt}^{\prime} 
g_{\phi \phi}^{\prime \prime}}{g_{tt}g_{\phi \phi}^{\prime}-g_{tt}^{\prime} g_{\phi \phi}} 
+\frac{2 g_{tt}^{\prime} g_{\phi \phi}^{\prime}}{g_{\phi \phi}g_{tt}}, 
\label{segundita}
\end{equation}

\noindent In the same way, using the explicit form of $E$ and $L$, 
(\ref{E2L2}), in (\ref{us}) 
expressions for the 4-velocities in terms of solely the metric components
are obtained

\begin{equation}
U^{\phi}=\sqrt{\frac{g_{t t}^{\prime}}
{g_{tt} g_{\phi\phi}^{\prime} - g_{tt}^{\prime} g_{\phi \phi}}}, \quad
U^t= -\sqrt{\frac{-g_{\phi \phi}^{\prime}}
{g_{tt} g_{\phi\phi}^{\prime} - g_{tt}^{\prime} g_{\phi \phi} }}.
\label{velocidades}
\end{equation}

\noindent 
and the angular velocity of the particles in these circular paths becomes
$\Omega = \sqrt{-g_{tt}^{\prime}/g_{\phi \phi}^{\prime}}$.
Since $b_{+}= - b_{-}$, the redshift $z_1=z_{red}$ and blueshift $z_2=z_{blue}$
are equal but with opposite sign: $z_1= -z_2$, the explicit expression is

\begin{equation}
z_1=\frac{-U^t_e U^{\phi}_d b_{d_{+}}+U^t_d U^{\phi}_e
  b_{e_{+}}}{U^t_d(U^t_d+U^{\phi}_db_{d_{+}})}.
\label{zfinal}
\end{equation}

\noindent
Moreover, if the detector is located far away from the compact object
$r_d \to \infty$, and as we mentioned before, 
$U_d^{\mu} \to (1,0,0,0)$; hence, (\ref{zfinal}) becomes

\begin{equation}
z_1 = U^{\phi}_e b_{e_{+}}= 
\sqrt{\frac{-g_{\phi \phi}g_{tt}^{\prime}}{g_{tt} (g_{tt} g_{\phi\phi}^{\prime} - 
g_{tt}^{\prime} g_{\phi \phi} )}}.
\label{zfinalFB}
\end{equation}

\section{Dilatonic Objects}

Since Dilaton and Einstein-Maxwell fields appear in the low energy
limit of string theory and as a result of dimensional reduction of
five-dimensional Kaluza-Klein theory, the four-dimensional effective
action of these theories can be written in the following form
\cite{horne} 

\begin{equation}
S= \int d^4x \sqrt{-g} \left [ -R + 2 (\nabla \Phi)^2 +
e^{-2 \alpha \Phi} F_{\mu \nu} F^{\mu \nu} \right ]
\label{accion}
\end{equation}

\noindent where $g=det(g_{\mu \nu})$, $R$ is the Ricci scalar, 
$\Phi$ the dilaton field, $F_{\mu \nu}$ the Faraday electromagnetic 
tensor and $\alpha \ge 0$ the dilaton coupling constant whose 
value determines the special theories contained in (\ref{accion}).
While $\alpha = \sqrt{3}$ leads to the Kaluza-Klein field equations
obtained by dimensional reduction from the five-dimensional Einstein
vacuum equations, $\alpha=1$ leads to the low energy 
limit of string theory, and $\alpha=0$
leads to the Einstein-Maxwell theory minimally coupled to the dilaton
scalar field. The field equations obtained from the action (\ref{accion})
read

\begin{equation}
\left ( e^{-2\alpha \Phi} F^{\mu \nu} \right )_{;\mu}=0,
\label{eqn1}
\end{equation}

\begin{equation}
\Phi^{;\mu}_{\mu} + \frac{\alpha}{2} e^{-2\alpha \Phi} F_{\mu \nu}F^{\mu \nu}=0,
\label{eqn2}
\end{equation}

\begin{equation}
R_{\mu \nu} = 2 \Phi_{,\mu} \Phi_{,\nu} + 2 e^{-2\alpha \Phi} 
\left ( F_{\mu \sigma}F_{\nu}^{\sigma}-\frac{g_{\mu \nu}}{4} 
F_{\beta \gamma}F^{\beta \gamma} \right )
\label{eqn3}
\end{equation}

\noindent where a comma means partial differentiation and a semicolon 
represents covariant derivative with respect to $g_{\mu \nu}$. 

\subsection{Generalized Chatterjee space-time}

The generalized Chatterjee space-time was found some years ago 
\cite{MaciasMatos}, and takes the form 

\begin{equation}
ds^2 = \frac{dr^2}{(1-\frac{2M}{r})^{\delta}} + 
(1-\frac{2M}{r})^{1-\delta} r^2 d\Omega^2
- (1-\frac{2M}{r})^{\delta} dt^2
\label{MM}
\end{equation}

\noindent where as usual $d\Omega^2=d\theta^2+\sin^2{\theta} d\phi^2$ and
$\delta$ is a free parameter. It possesses a dilaton scalar field yet 
without electromagnetic fields. The dilaton field reads

\begin{equation}
e^{-2\alpha \Phi} = \frac{\kappa_0^2}{4} \left ( 1- \frac{2M}{r} \right )^{-\alpha \sqrt{|1-\delta^2|}}
\label{dilaton}
\end{equation}

\noindent It is apparent that for $\delta=1$ the dilaton field 
becomes constant and (\ref{MM}) becomes the Schwarzschild black hole 
with its event horizon at $r=2M$. For $\delta=1/2$, the metric
components are real provided that $r>2M$. We will restrict ourselves
to the region $r>2M$ henceforth. 

The solution (\ref{MM}) includes the following three known 
cases (see the details in \cite{MaciasMatos}):

\begin{enumerate}
\item $\delta=1$ corresponds to the Schwarzschild black hole.

\item $\delta=1/2$ and $\alpha=\sqrt{3}$, the solution (\ref{MM}) reduces
to the Kaluza-Klein soliton solution which corresponds to the
Chatterjee solution. 

\item $\delta=2$ and $\alpha$ arbitrary represents a black hole in the 
dilaton gravity theory framework. 

\end{enumerate}

\subsubsection{Case $\delta=1$}

In \cite{us} the relationship (\ref{zfinalFB}) for a Schwarzschild 
black hole ($\theta=\pi/2$) turned out to be 

\begin{equation}
M= r_c \mathcal{F}(z) \quad \text{where}
\quad \mathcal{F}_{\pm}(z)=\frac{1+5z^2\pm \sqrt{1+10z^2+z^4}}{12z^2}.
\label{Mrz}
\end{equation}

\noindent being $M$ the mass parameter, $r_c$ the radius 
of a massive particle's circular orbit 
that emits light and $z$ its redshift. These
circular orbits are stable as long as 
$V_{eff}^{\prime \prime}>0$. Since 

\begin{equation}
V_{eff}^{\prime \prime}= \frac{2M(r_c-6M)}{r_c^2(r_c-2M)(r_c-3M)},
\label{V2explicita}
\end{equation}

\noindent is positive provided that $r_c > 6 M$; 
then $\frac{r_c}{M}= \mathcal{F}^{-1} > 6$ which 
is fulfilled if and only if $|z| < 1/\sqrt{2}$ 
and solely for the minus sign $\mathcal{F}_{-}(z)$. 
Therefore, a measurement of the redshift $z$ of light 
emitted by a particle that follows
a circular orbit of radius $r_c$ in the equatorial plane around a
Schwarzschild black hole will have a mass parameter determined by 
$M = r_c \mathcal{F}_{-}(z)$, and $z$ must be such that $|z|< 1/\sqrt{2}$.

The energy, angular momentum, velocities $U^t$, $U^{\phi}$ and 
the angular velocity of 
the emitter, can be computed from (\ref{E2L2}), 
(\ref{velocidades}) and written as function
of the measurable redshift $z$ and radius $r_c$ of the circular 
orbit of the photon source \cite{us}. Since $|z|<1/\sqrt{2}$,
all these five quantities $E^2,L^2,U^{t},U^{\phi},\Omega$
are bounded by $z$ given a specific circular orbit of radius $r_c$.

\subsubsection{Cases $\delta \neq 1$}

For the Schwzarschild black hole, the redshift is bounded as 
$|z|<1/\sqrt{2}$. We will determine now
whether $z$ is bounded for $\delta \neq 1$ and the dependance
of the mass parameter $M$ in terms of the frequency shift of photons
emitted by particles traveling in circular orbits of radius $r_c$, that is,
 $M=M(r_c,z)$ will be found for any given $\delta \neq 1$. 
To that end, we start with the relationship (\ref{zfinalFB}) 
which for this metric becomes

\begin{equation}
F(M;r,z)=z^2 \left [ r-M(1+2\delta) \right ] \left ( 1- \frac{2M}{r}
\right )^{\delta} \ - \delta M =0 . 
\label{z2chatterjee}
\end{equation}

\noindent There is no analytic solution for this equation for arbitrary
$\delta$, its roots have to be found numerically. 
Circular orbits on the equatorial plane exist provided that the
following two conditions 

\begin{eqnarray}
E^2 &=& \left ( 1-\frac{2M}{r} \right )^{\delta} 
\frac{r-M(1+\delta)}{r-M(1+2\delta)} >0 \nonumber \\
L^2 &=& \left ( 1-\frac{2M}{r} \right )^{-\delta} 
\frac{M r \delta (r-2M)}{r-M(1+2\delta)} >0
\label{E2L2Chatter}
\end{eqnarray}

\noindent are simultaneously fulfilled. Since $r>2M$, $L^2$ is
positive as long as $r-M(1+2\delta)>0$, the fulfillment of this
inequality also guarantees that $E^2>0$. Stability of these circular orbits
requires 

\begin{equation}
V''(r)=\frac{2M \delta \left [ (2+6\delta+4\delta^2)M^2 + r^2 - 2Mr (1+ 3 \delta) \right ] }{r^2 (r-2M)^2 (r-M(1+2\delta))}
\label{V2charttejee}
\end{equation}

\noindent be positive. For $\delta < 1/\sqrt{5}$, it turns that
$V''>0$ for all $r>2M$ and $r>M(1+2\delta)$ (which are the conditions to have
circular orbits). For $\delta \ge 1/\sqrt{5}$, (\ref{V2charttejee}) 
can be written as

\begin{equation}
V''= \frac{2M\delta (r-M \Lambda_{+})(r-M\Lambda_{-})}{r^2(r-2M)^2(r-M(1+2\delta))},
\label{VppFactor}
\end{equation}

\noindent where $\Lambda_{\pm}=1+3\delta \pm \sqrt{5\delta^2-1}$.
From this expression, it is apparent that circular orbits will be stable
for either $\Lambda_{+} < r/M$ or $r/M <\Lambda_{-}$. To perform the
analysis, we proceed in the following manner: for a fixed value of 
$\delta$, a domain $\mathcal{D}_{r,z}= (r_{min},r_{max}) \times (z_{min},z_{max})$
is set and a search for positive roots of (\ref{z2chatterjee})
is carried out. We numerically find these roots using a hybrid 
algorithm, a combination of bisection and Newton-Raphson 
methods \cite{NR}.
At some points of the domain $\mathcal{D}_{r,z}$
there are more than one positive root of (\ref{z2chatterjee}). 
We then test whether 
the conditions (\ref{E2L2Chatter}) and $V'' >0$ for circular and stable orbits
are simultaneously satisfied. Those
roots of (\ref{z2chatterjee}) at a point $q \in \mathcal{D}_{r,z}$ 
that do not fulfill the conditions for circular and stable orbits are 
discarded. What we have found is that, not for every single point 
$q \in \mathcal{D}_{r,z}$ 
there is a root of $F(M;r,z)=0$ that
leads us to a circular stable orbit of radius $r_c$ 
followed by a photon emitter particle; only in a
subset $\mathcal{D} \subset \mathcal{D}_{r,z}$ does such a mass parameter exist.
Moreover, in all the surveys we have done on domains
with different sizes, in every point $q \in \mathcal{D}_{r,z}$ 
the parameter $M$ attained is unique.
We work with geometrized units ($G=c=1$) and we scale $M$ and $r$ by a 
multiple of the solar mass $p M_{\odot}$.

\begin{figure}[htp]
  \resizebox{70mm}{!}{\includegraphics{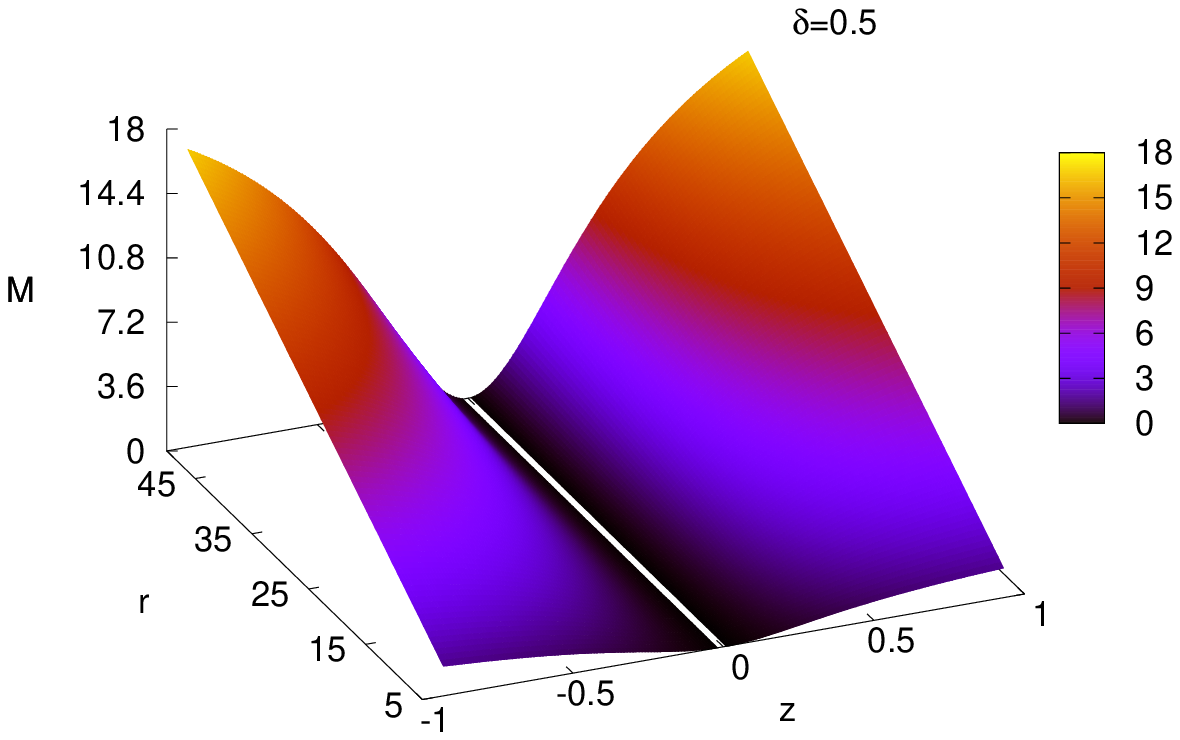}}\\
  \resizebox{70mm}{!}{\includegraphics{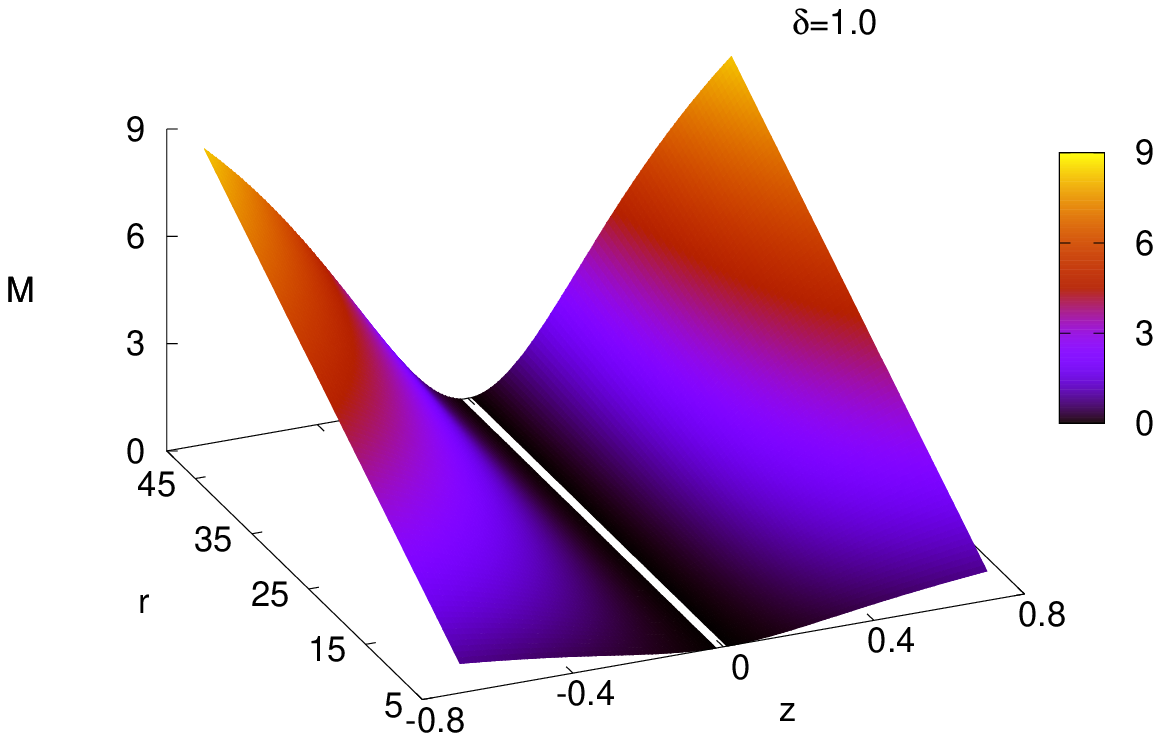}}\\
   \resizebox{70mm}{!}{\includegraphics{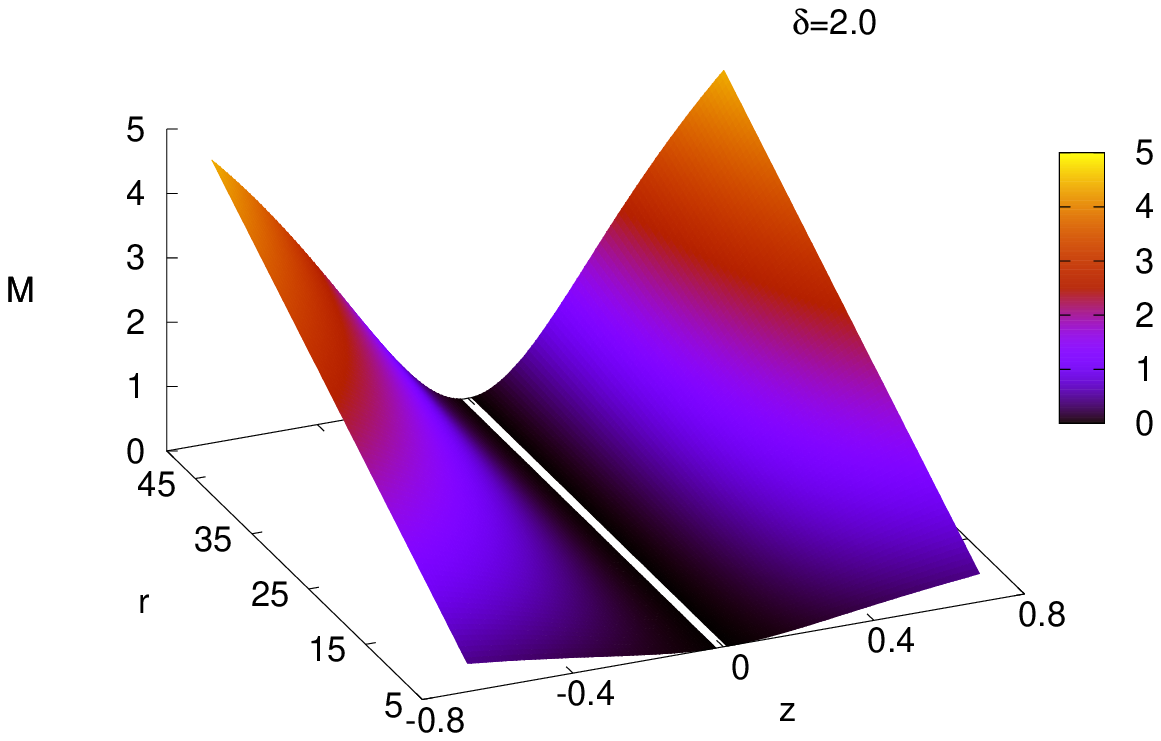}}\\ 
 \caption{Mass parameters in terms of redshift ($z>0$) or 
blueshift ($z<0$) of photons emitted by particles 
in circular orbits of radius $r$ around
Chatterjee space-time. The upper graph corresponds to 
the cases of Kaluza-Klein soliton ($\delta=1/2$), the middle
graph corresponds to Schwarzschild black hole ($\delta=1$)
and lower graph corresponds to a black hole in the dilaton gravity framework
($\delta=2$). $M$ and $r$ are in geometrized units and scaled by 
$p M_{\odot}$ where $p$ is an arbitrary proportionality factor.}
 \label{masses}
\end{figure}

We show the plots of $M=M(z,r_c)$ in figure \ref{masses} 
for the three special cases: Kaluza-Klein soliton
($\delta=0$.$5$), Schwarzschild black hole ($\delta=1$) and Dilaton
black hole ($\delta=2$). It can be seen that as $\delta$ increases 
its values, the mass obtained decreases and it is symmetric with respect to 
the frequency shift $z$ ($z_{red}>0$ and $z_{blue}<0$). Given a set of $N$ 
pairs $\{ r_c, z\}_i$ of observed redshifts (blueshifts) $z$ of emitters 
traveling around a Chatterjee object along circular orbits of radii $r_c$, 
a Bayesian statistical analysis could be carried out in order to estimate
the black hole mass parameter.
 
\begin{figure}[htp]
  \resizebox{60mm}{!}{\includegraphics{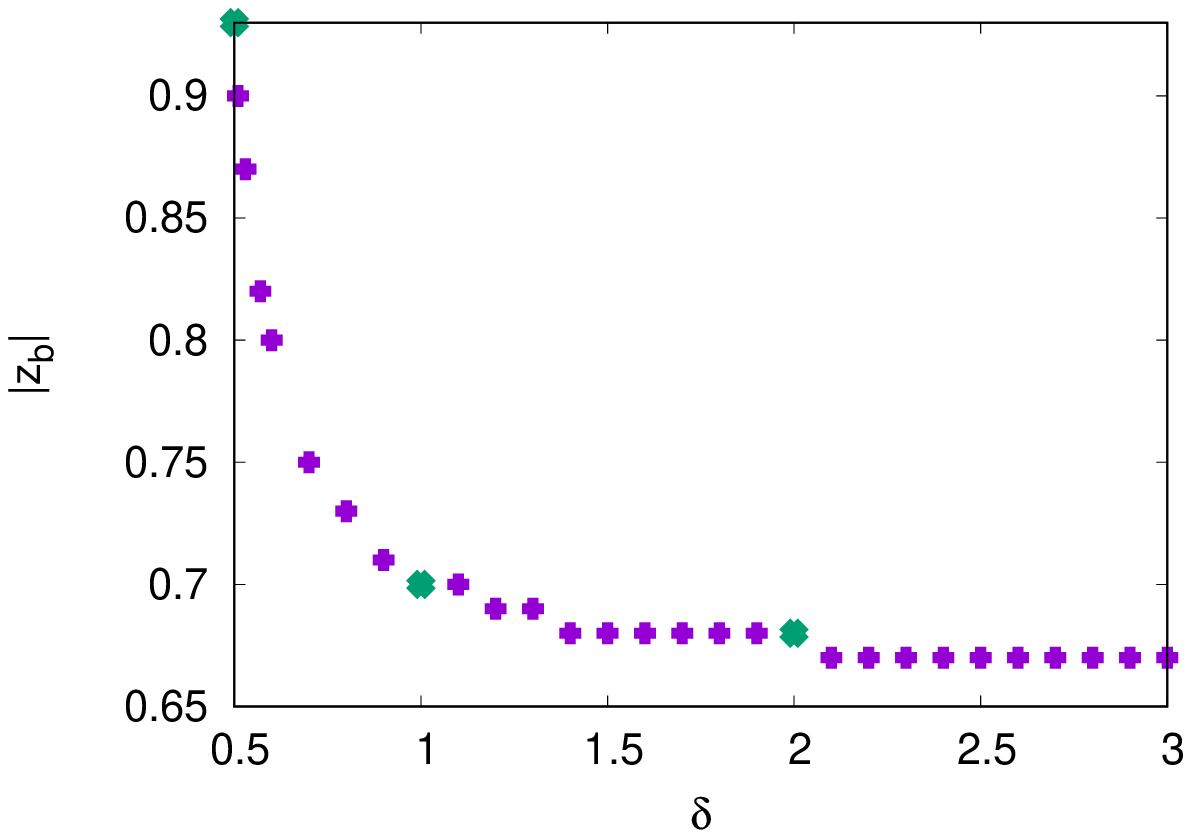}}\\
  \resizebox{60mm}{!}{\includegraphics{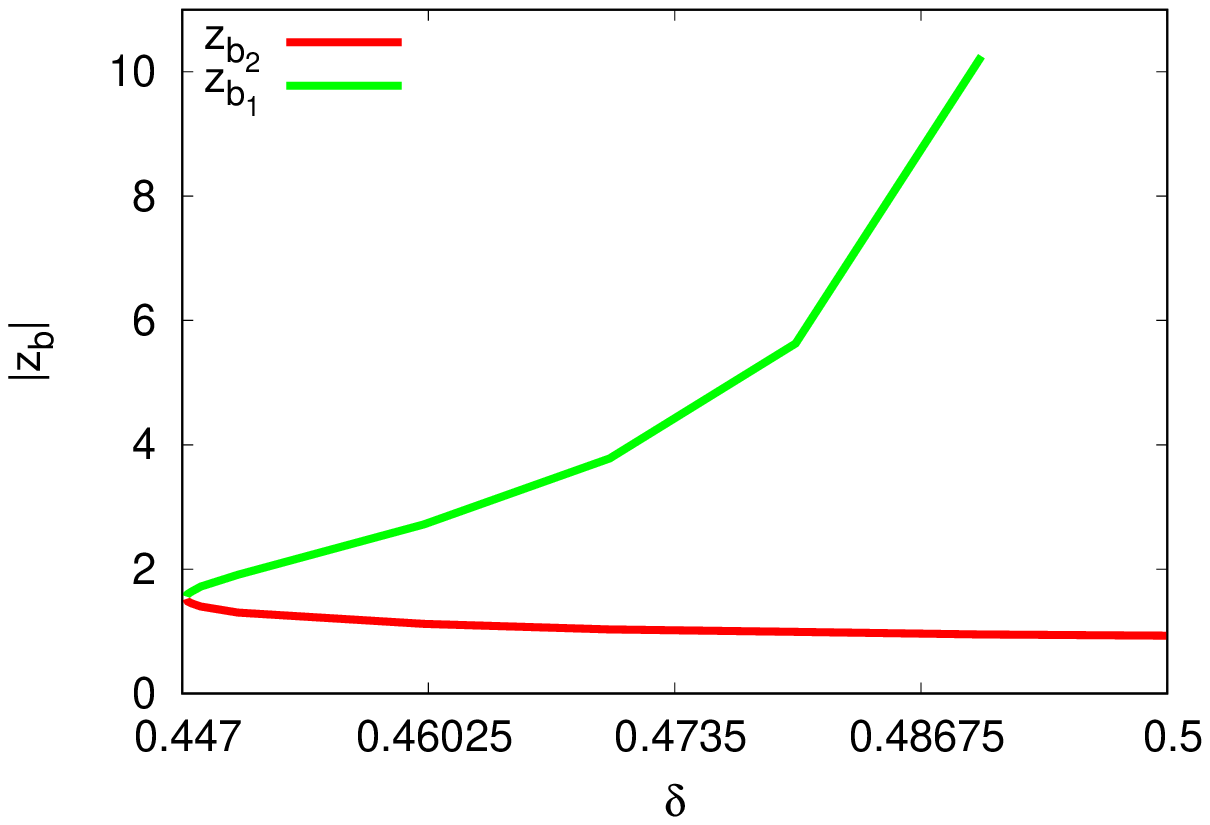}}\\
 \caption{ The upper plot shows the value $z_b=z_b(\delta)$
such that, for $|z|<z_b$ there exists a 
unique value of the mass parameter $M(r,z)$. 
Asterisks (in blue) represent the
   three known cases: Kaluza-Klein solution ($\delta=1/2$), Schwarzschild
   black hole ($\delta=1$) and dilaton black hole ($\delta=2$).
In the lower plot, the two regions $|z| > z_{b2}$ and $|z|<z_{b1}$ are
physically aceptable. }
  \label{deltavsz}
\end{figure}

In regard to the frequency shifts bounds, for $\delta > 1/2$, for each 
$\delta$ there is a $z_b$ such that for $|z|< z_b$, there exists a 
unique value of the mass parameter $M(r_c,z)$, and $z_b(\delta)$ 
does not depend on the radii $r_c$ of the circular orbits of photon emitters. 
$z_b=z_b(\delta)$ is shown in the upper plot of figure 
\ref{deltavsz}, asterisks (in blue) represent the
three known cases: Kaluza-Klein solution ($\delta=1/2$), 
Schwarzschild black hole ($\delta=1$) and dilaton black hole 
($\delta=2$). For $\delta$'s in a small vicinity of unity, the bound $z_b$
varies little around $1/\sqrt{2}$ which is the Schwarzschild frecuency shift
bound. For $\delta \in (1$.$7,3)$, $|z_b - 1/\sqrt{2}|<0$.$04$.
Nonetheless, for $\delta \in (1/\sqrt{5},1/2)$, there is a 
significant departure from that Schwarzschild bound; furthermore, 
there are two frequency shifts regions where there is a positive root
$M(r_c,z)$ of (\ref{z2chatterjee}) corresponding to circular and stable orbits
of photon emitters. The lower plot in figure \ref{deltavsz} shows 
these two regions: $|z| > z_{b2}$ and $|z|<z_{b1}$ are
physically aceptable. For $0 < \delta < 1/\sqrt{5}$ the stability condition 
$V''>0$ always holds and there is no bound for $z$.

\begin{figure}[htp]
  \resizebox{55mm}{!}{\includegraphics{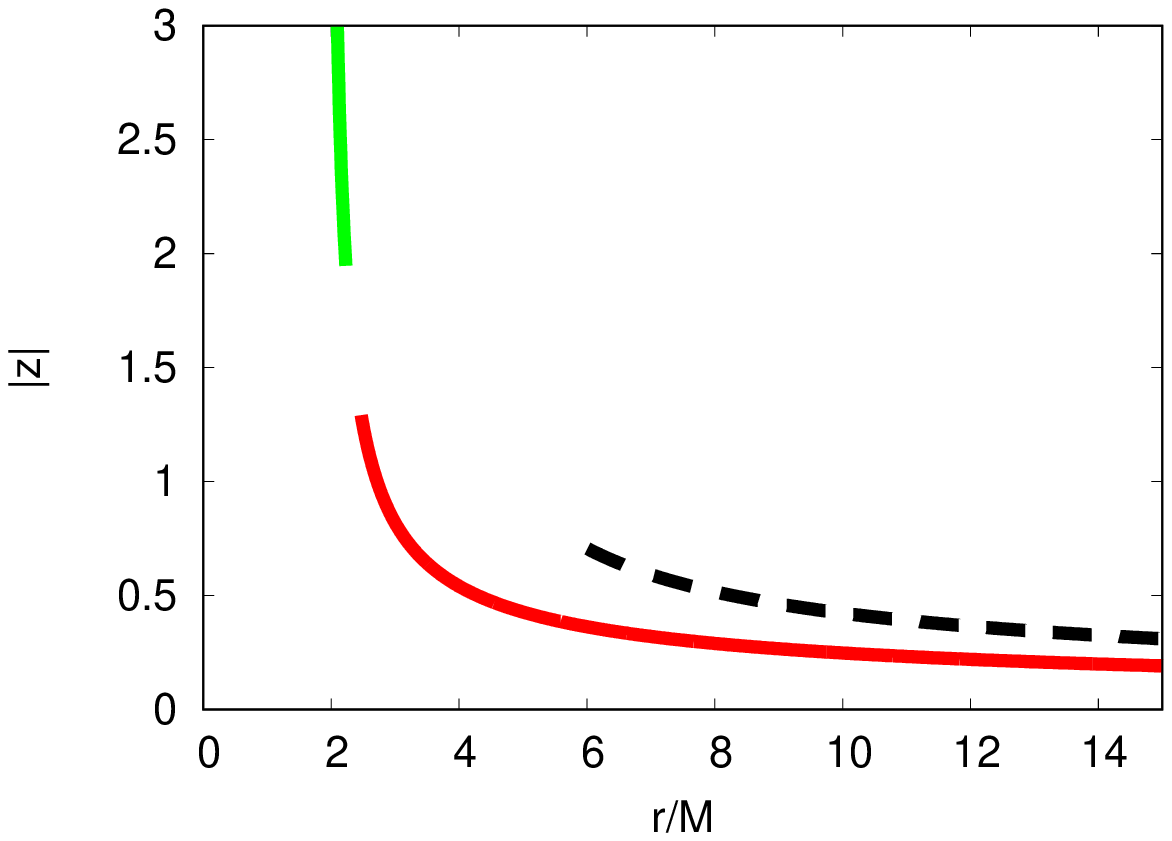}} \vspace{0.7cm}\\
  \resizebox{55mm}{!}{\includegraphics{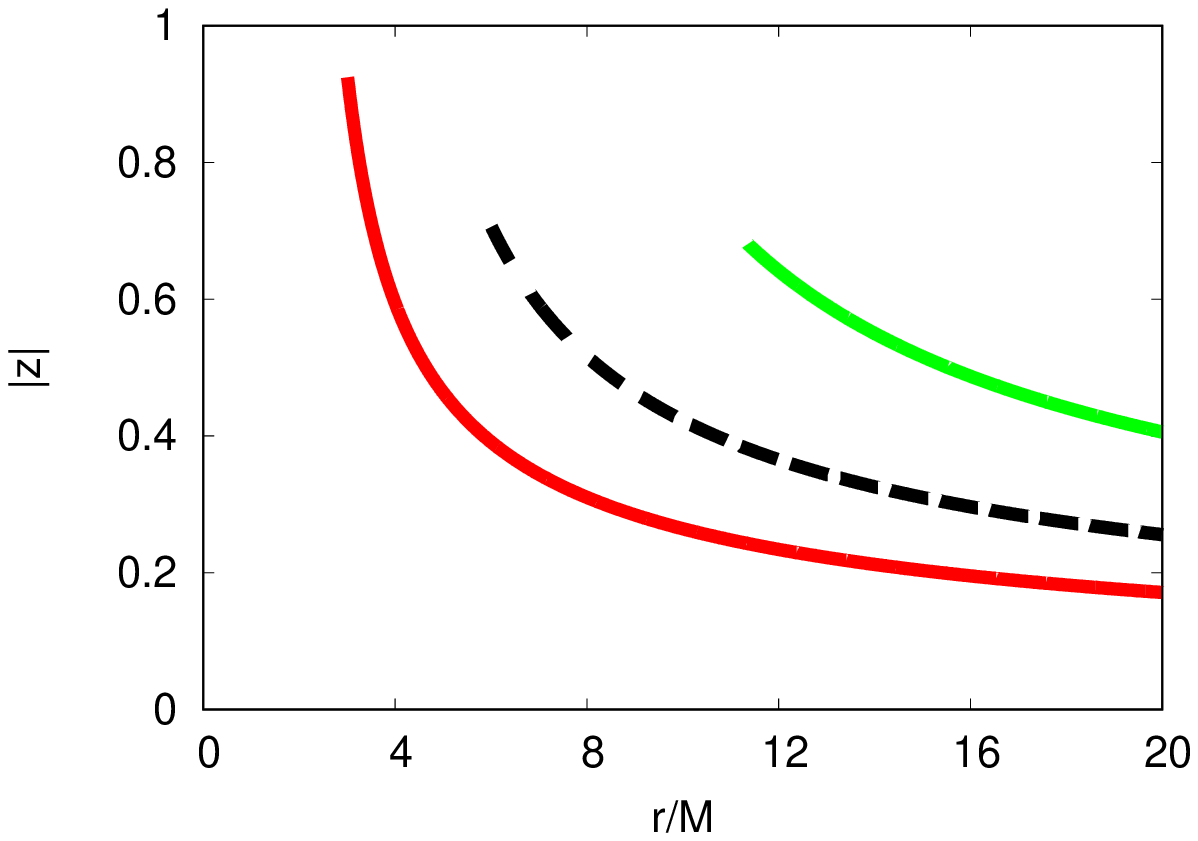}}\\
 \caption{The upper plot ($\delta=0$.$45$) has a gap corresponding
to $|z| > z_{b2}$ (green curve) and $|z|<z_{b1}$ (red curve). 
The lower plot shows $|z|=|z|(r/M)$ for $\delta=0$.$5$ 
(red curve) and $\delta=2$ (green curve). 
In both plots, the dashed black curves correspond to the 
Schwarszchild black hole. }
  \label{zChatt}
\end{figure}

The upper plot in figuer \ref{zChatt} shows the frequency shift $|z|$ as 
function of $r/M$ for $\delta=0$.$45$. There is a discontinuity in the 
frequency shift values physically acceptable corresponding to
$|z| > z_{b2}$ (green curve) and $|z|<z_{b1}$ (red curve). 
The lower plot shows  $|z|=|z|(r/M)$ for $\delta=0$.$5$ 
(red curve) and $\delta=2$ (green curve). 
In the upper and lower plots, the dashed black curves 
correspond to the Schwarszchild black hole.
For $\delta$'s not close to unity, there is a evident departure 
from the Schwarzschild curve that might allow
us to distinguish (if observational data were available)
whether a Schwarzschild or Chaterjee space time is in the center of a galaxy.
 
\subsection{Gibbons-Maeda space-time}

The Gibbons-Maeda metric could represent a charged static black hole
with a scalar field \cite{GM}. This metric reads

\begin{equation}
ds^2=-( 1-\frac{2M}{r}) dt^2 + ( 1-\frac{2M}{r})^{-1}
dr^2 + r (r- \frac{Q^2}{M}) d\Omega^2
\label{GM}
\end{equation}

\noindent
For $Q=0$ (\ref{GM}) becomes the Schwarzschild black hole. We shall work in the
region $r>2M$. Particles will follow circular and equatorial orbits 
in Gibbons-Maeda space-time provided that

\begin{eqnarray}
L^2 &=& \frac{2r(Q^2-Mr)^2}{2M(2Q^2+r^2)-(6M^2+Q^2)r} > 0 \nonumber \\
E^2 &=& \frac{(r-2M)^2(-Q^2+2Mr)}{2M(2Q^2+r^2)-(6M^2+Q^2)r} > 0. 
\label{E2L2GM}
\end{eqnarray}

\noindent $L^2>0$ holds if and only if 

\begin{eqnarray}
P(r,M)&=&2M(2Q^2+r^2)-(6M^2+Q^2)r \nonumber \\
&=& 2M (r-r_{-})(r-r_{+})>0
\label{PrM}
\end{eqnarray}

\noindent  where

\begin{equation}
r_{\pm}=\frac{6M^2+Q^2\pm 6\sqrt{(M^2-Q^2/2)(M^2-Q^2/18)}}{4M} 
\label{rmasmenos}
\end{equation}

In order for $r_{\pm}$ to be real, $2M^2>Q^2$ must be required.  
Since for Schwarzschild black hole, circular orbits exist solely for 
$r>3M$ and $r_{-} \to 0$ and $r_{+} \to 3M$ as $Q \to 0$, we work in 
the region $r>r_{+}$. It turns out that $r>r_{+}$ and $2M^2>Q^2$ imply
the fullfilment of $E^2>0$. Circular orbits are stable as long as the condition

\begin{equation}
V''=\frac{-4M^2(2Q^4-6MrQ^2+M(6M-r)r^2)}{P(r,M) r^2 (Mr-Q^2)(r-2M)} > 0, 
\label{VppGM}
\end{equation}

\noindent holds and this is so as long as
$Mr^3-6M^2r^2+6MQ^2r-2Q^4>0$. The relation that connects $M,r,Q^2$ and $z$ 
comes from (\ref{zfinalFB}) and it reads

\begin{equation}
z^2=\frac{2Mr(Mr-Q^2)}{(r-2M) \left [ 2M(2Q^2+r^2)-(6M^2+Q^2)r \right ] }
\label{z2Maeda}
\end{equation}

In order to find $M=M(r_c,z;Q^2)$ one has to solve this cubic equation for $M$.
Equation (\ref{z2Maeda}) have either three real roots or one real plus 
two imaginary roots. We keep only the real positive roots $M=M(r_c,z;Q^2)$ that 
fulfill the conditions for circular and stable orbits and it turns out
that, from the analytic Cardan's formula, the only one that satisfies
those conditions is

\begin{equation}
M(r_c,z,Q^2)=\left [ 2  \sqrt{\frac{-p}{3}} \cos{(\frac{\phi}{3}+\frac{4 \pi}{3})}- \frac{a}{3} \right ] \quad \mbox{where} \nonumber
\end{equation}

\begin{eqnarray}
p &=& b - \frac{a^2}{3}, \quad q = c-\frac{ab}{3}+\frac{2a^3}{27}, 
\quad
\phi = \cos^{-1}\left [ \frac{\sqrt{27} q}{2p\sqrt{-p}} \right ] \nonumber \\
a &=& -\left ( \frac{2Q^2}{3r_c} + \frac{5z^2+1}{6z^2}r_c \right ), \quad
c = - \frac{r_cQ^2}{12} \nonumber \\
b &=& \left ( Q^2 \frac{3z^2+1}{6z^2} + \frac{r_c^2}{6} \right ). 
\label{pqphi}
\end{eqnarray}

\begin{figure}[htp]
\centerline{ \epsfysize=6.5cm \epsfbox{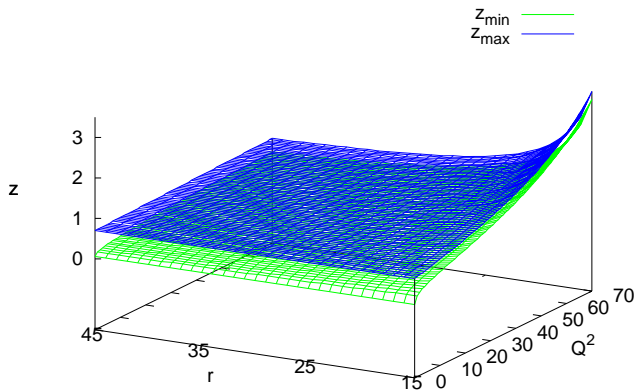}}
 \caption{The two redshift surfaces $z_{min}$ and $z_{max}$
as function of the radius $r$ of circular orbits followed by photon
emitters around a Gibbson-Maeda spacetime and its charge
parameter $Q^2$ are shown. Only for redshifts $z \in (z_{min},z_{max})$
the corresponding values $M=M(r,z,Q^2)$ are acceptable.
$M$, $Q$, and $r$ are in geometrized units and scaled by $pM_{\odot}$ 
where $p$ is an arbitrary factor of proportionality .}
  \label{Cotas}
\end{figure}

This analysis is performed numerically in the following
manner: we set a domain $\mathcal{D}=(Q_1,Q_2)\times (r_1,r_2) \times (z1,z2)$,
for each $q \in \mathcal{D}$ the roots of (\ref{z2Maeda}) are
found. With these roots at hand, we check whether the following conditions
are all satisfied: (i) $r>2M$, (ii) $r > r_{+}$, (iii) $2 M^2 > Q^2$ and 
(iv) $M r^3 - rM^2 r^2 + 6 MQ^2r-2Q^4 > 0$. The second and third inequalities
guarantee that indeed, we have circular and equatorial orbits; the fourth 
inequality guarantees stability of these orbits. As mentioned, it
turns out that the only root satisfying all the conditions has the form 
given by (\ref{pqphi}). For given values of $r_c$ and $Q^2$, 
we search for the minimum and maximum value of $z$ for which these four 
conditions simultaneously hold, this process yields bounds of the frequency
shifts. For $Q= 0$, the result for Schwarzschild ($|z| < z_b= 1/\sqrt{2}$) 
is recovered as it should be. Figure \ref{Cotas} shows the surfaces 
$z_{min}=z_{min}(r_c,Q^2)$ and $z_{max}=z_{max}(r_c,Q^2)$. Only for 
frequency shifts $z$ such that $|z| \in (z_{min},z_{max})$,
the corresponding values for the mass parameter $M=M(r_c,z,Q)$ 
are acceptable. We observe that the larger $Q^2$ is, the narrower the gap
between $z_{min}$ and $z_{max}$ becomes. For large $Q^2$'s and small $r_c$'s,
$z_{min}$ and $z_{max}$ sharply increase. 

\begin{figure}[htp]
  \resizebox{75mm}{!}{\includegraphics{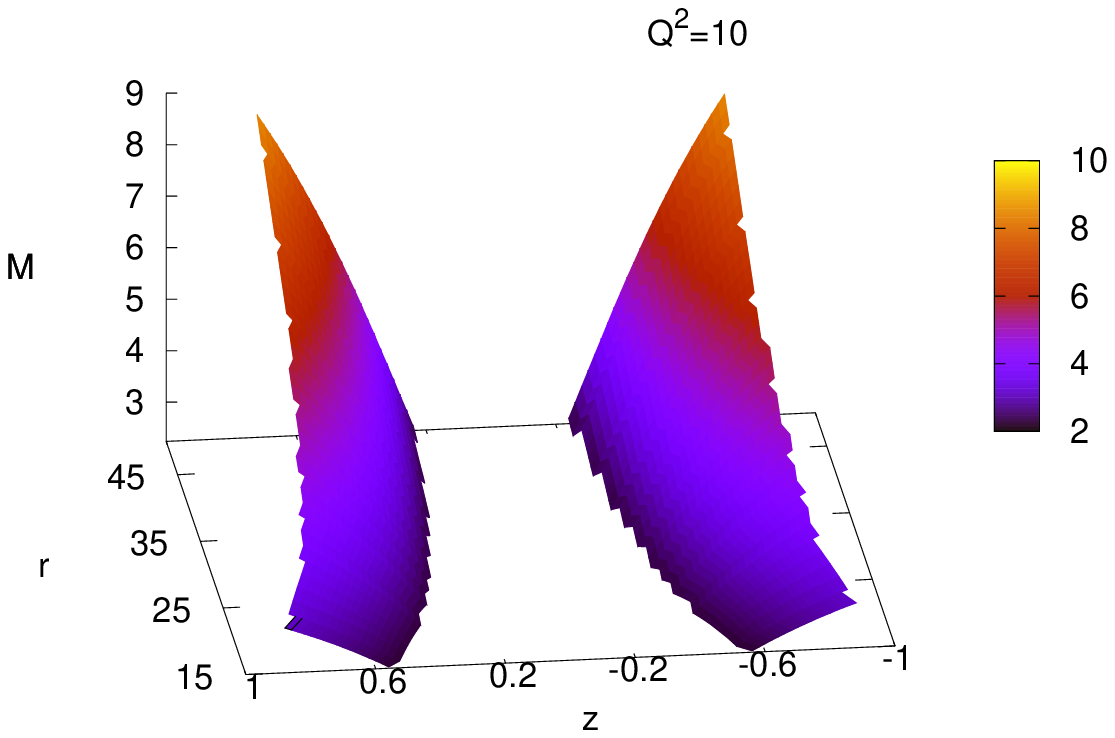}}\\
  \resizebox{75mm}{!}{\includegraphics{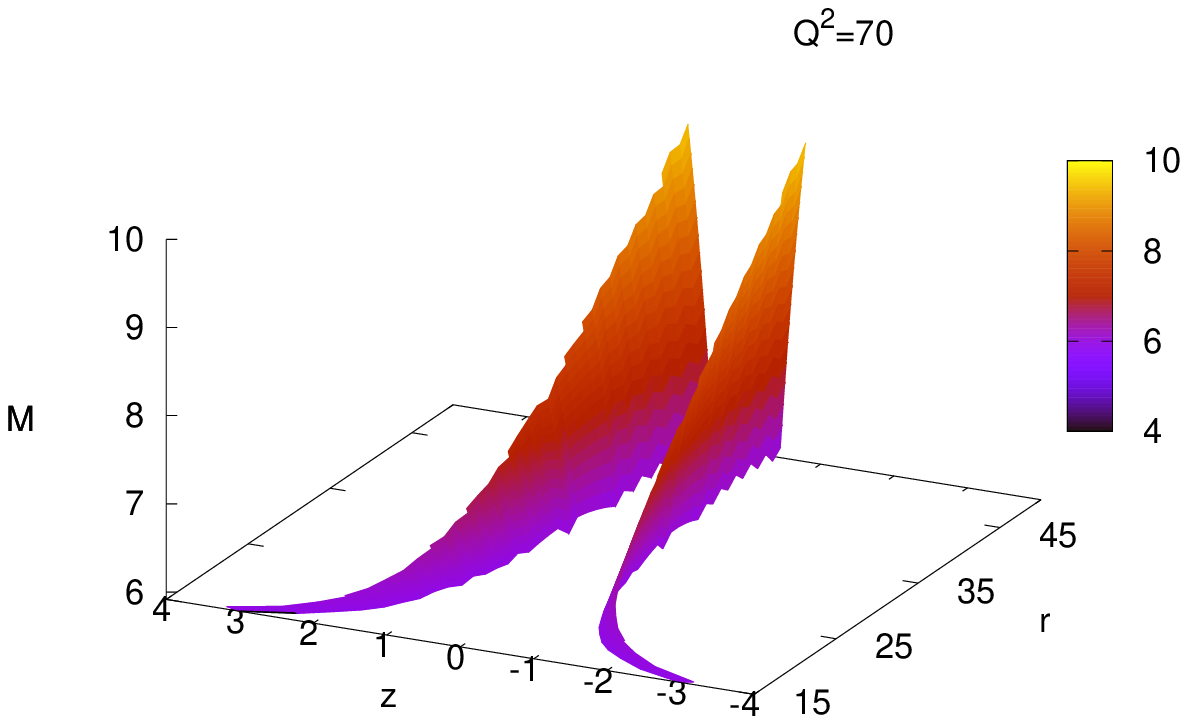}}\\
 \caption{Mass parameters in terms of redshift ($z>0$) and
blueshift ($z<0$) of photons emitted by particles 
in circular orbits of radius $r$ around
Gibbons-Maeda space-time. The upper plot shows $M(r,z,Q^2=10)$ and the
lower plot $M(r,z,Q^2=70)$. $M$, $r$ and $Q$ 
are in geometrized units and scaled by $p M_{\odot}$ where $p$ is an 
arbitrary proportionality factor.}
  \label{MGM}
\end{figure}

Due to the condition (iii) $2M^2 > Q^2$, as the charge $Q^2$ increases, 
the corresponding acceptable value $M(r_c,z,Q^2)$ also increases. The upper
plot of figure \ref{MGM}, shows $M=M(r_c,z,Q^2=10)$ and $M>\sqrt{5}$ as it
should be according condition (iii), the lower plot shows
$M=M(r_c,z,Q^2=70)$ and $M>\sqrt{35}$.
$M$, $r_c$ and $Q$ are in geometrized units and scaled by $p M_{\odot}$ 
where $p$ is an arbitrary proportionality factor. For $Q=0$, the result
of the Schwarzschild black hole shown in the middle plot of figure 
\ref{masses} with $z_b=1/\sqrt{2}$ is recovered.

\section{Final remarks}

There have been efforts to find astrophysical signatures of scalar fields.
In \cite{Matos-Brena} the following spacetime was considered

\begin{eqnarray}
ds^2 &=&-( 1-\frac{2M}{r} ) dt^2 + e^{2k_s} \frac{dr^2}{1-2M/r}
\nonumber \\
&& + r^2 \left ( e^{2k_s} d\theta^2 + \sin{\theta} d\phi^2 \right )
\label{MB}
\end{eqnarray}

\noindent where

\begin{eqnarray}
e^{2k_s}&=& \left ( 1 + \frac{M^2 \sin^2{\theta}}{r^2 (1-2M/r)} \right )^{-1/b^2}
\nonumber \\
\Phi&=&\frac{1}{2b} \ln \left( 1-\frac{2M}{r} \right ),
\label{PhiMB}
\end{eqnarray}

\noindent with $b$ an integration constant. As $b \to \infty$, the scalar field
vanishes and the Schwarzschild black hole is recovered. The metric 
(\ref{MB}) was employed to model the Sun and the deflection of light 
rays passing nearby was computed, the authors found that the current
observational errors for this effect set the limit $b > 0.02$. One of their
conclusions was that even if a scalar field of the form (\ref{PhiMB}) 
were actually present in the solar system, it cannot be presently detected.
If we apply HN formalism to the metric (\ref{MB}) considering equatorial
and circular orbits, then $M=M(r_c,z)$ and the bound for $z$ 
would be exactly the same as the one for the Schwarzschild black hole 
since the components $g_{rr}$ and $g_{\phi \phi}$
are just the same. As a matter of fact, any solution to the EMD theory that
for equatorial and circular orbits leaves $g_{rr}=(1-2M/r)^{-1}$ and
$g_{\phi \phi}=r^2$ would yield the same results as for that of the 
Schwarzschild metric.

From our analysis on the Chatterjee spacetime, we observed that, for 
$\delta$'s not close to unity, the departure of our results from those
of the Schwarzschild black hole is noticeable and becomes truely apparent 
for $\delta \in (1/\sqrt{5},1/2)$ where
there are two frequency shifts regions where there is a positive root
$M(r_c,z)$ of (\ref{z2chatterjee}) corresponding to circular and stable orbits
of photon emitters. For the Gibbons-Maeda metric, the bounds of frequency
shifts are two surfaces $z_{min}(r_c,Q^2)$ and $z_{max}(r_c,Q^2)$ whose gap
narrows as $Q^2$ increases and $r_c$ decreases their values, as shown in 
figure \ref{Cotas}. It would be interesting to apply this formalism to
rotating dilaton black holes \cite{horne} and contrast the results with those
found in \cite{us} for the Kerr black hole. \\

\noindent {\bf ACKNOWLEDGMENTS} \\

S.V-A acknowledges partial support from PRODEP, under 4025/2016RED project.
F.A and R. B. acknowledge partial support by CIC-UMSNH. 
The authors thank professor Ulises Nucamendi for useful comments and 
discussions on the results of this work.


\thebibliography{99}

\bibitem{evidence} M. B. Begelman, Evidence for black holes, Science
  {\bf 300}, 1898 (2003). Z. Q. Shen, K. Y.  Lo, M.-C. Liang,
  P. T. P.  Ho, and J.-H. Zhao, A size of $\approx$ 1 au for the radio
  source Sgr $A^*$ at the center of the Milky Way, Nature (London)
  {\bf 438}, 62 (2005). A. M. Ghez, S. Salim, N. N. Weinberg,
  J. R. Lu, T. Do,  J. K. Dunn, K. Matthews, M. R. Morris, S. Yelda, E.
  E. Becklin, T. Kremenek, M. Milosavljevic, and J. Naiman, Measuring
  distance and propereties of the Milky Way's central supermassive
  black hole with stellar orbits, Astrophys. J. {\bf 689}, 1044
  (2008). M. R. Morris, L. Meyer, and A. M. Ghez, Galactic center
  research: Manifestations of the central black hole,
  Res. Astron. Astrophys. {\bf 12}, 995 (2012)

\bibitem{ulises} Alfredo Herrera and Ulises Nucamendi,
 Kerr black hole parameters in terms of the redshift/blueshift of photons
emitted by geodesic particles, Phys. Rev. D {\bf 92}, 045024 (2015).

\bibitem{us} Ricardo Becerril, Susana Valdez-Alvarado, Ulises Nucamendi,
Obtaining mass parameters of compact objects from redshifts and blueshifts
emitted by geodesic particles around them, Phys. Rev. D {\bf 94},
124024 (2016). 

\bibitem{estimates}
B. Aschenbach, N. Grosso, D. Porquet and P. Predehl, 
X-ray flares reveal mass and angular momentum of the Galactic Center 
black hole, A \& A {\bf 417}, 71–794, 8 (2004).

\bibitem{LS} T. Matos, F.S. Guzman  and L. Ure\~na. Scalar Fields as Dark 
Matter in the Universe, Class. Quantum Grav. {\bf 17}, (2000)
1707. Mikel Susperregi, Dark matter and dark energy from an
inhomogeneous dilaton, Phys. Rev. D {\bf 68}, 123509 (2003). 

\bibitem{RC} T. Matos and F.S. Guzman. Scalar fields as dark matter in 
spiral galaxies, Class. Quantum Grav. {\bf 17}, L9 (2000).

\bibitem{Lens} T. Matos and R. Becerril, An axially symmetric scalar field
as a gravitational lens, Class. Quantum Grav. {\bf 18}, 2015 (2001).

\bibitem{Dynamics} Eric W. Hirschmann, Luis Lehner, Steven L. Liebling,
and Carlos Palenzuela, Black Hole Dynamics in Einstein-Maxwell-Dilaton Theory.
 arXiv:1706.09875, (2017).

\bibitem{MaciasMatos} T. Matos and A. Mac\'ias. Black holes from generalized
Chatterjee solutions in dilaton gravity, Modern Physics Letters A {\bf 9},
3707 (1994).

\bibitem{GM} G.W.Gibbons and Kei-ichi Maeda.
Black holes and membranes in higher-dimensional theories with dilaton fields,
Nuclear Physics B {\bf 298}, 741 (1988).
David Garfinkle, Gary T. Horowitz and Andrew Strominger,
Charged black holes in string theory, Phys. Rev. D {\bf 43}, 3140 (1991).

\bibitem{NR} William H. Press, Saul A. Teukolsky, William T. Vetterling and Brian P. Flannery, Numerical Recipes. Cambridge University Press.

\bibitem{Matos-Brena} T. Matos and Hugo V. Brena, Possible astrophysical signatures of scalar fields, Class. Quantum Grav. {\bf 17}, 1455 (2000).

\bibitem{horne} James H. Horne and Gary T. Horowitz, Rotating dilaton black holes. Phys. Rev. D {\bf 46}, 1340 (1992).
\end{document}